%% file: main.tex
\documentclass[sigconf]{acmart}

\AtBeginDocument{%
  }

\setcopyright{acmlicensed}
\copyrightyear{2018}
\acmYear{2018}
\acmDOI{XXXXXXX.XXXXXXX}

\acmConference[Conference acronym 'XX]{Make sure to enter the correct
  conference title from your rights confirmation emai}{June 03--05,
  2018}{Woodstock, NY}

\acmISBN{978-1-4503-XXXX-X/18/06}

\usepackage{acronym}
\usepackage{multirow}
\usepackage{caption}
\usepackage{siunitx}
\usepackage{subcaption}
\usepackage{xspace}
\usepackage{xcolor}
\usepackage{enumitem}
\usepackage{array}
\usepackage{tcolorbox}
\usepackage{subcaption}
\usepackage{tcolorbox}
\usepackage{amsmath}
\usepackage{booktabs}
\usepackage{ulem}
\usepackage{amsfonts}

\usepackage{xcolor}

\input{definitions}

\newcommand{\hypbox}[2]{%
\begin{tcolorbox}[colback=white!98!black,colframe=white!30!black,boxsep=1.1pt,top=6.75pt]%
\vspace{1.75pt}%
\textbf{#1}\\[-0.575em]
\noindent\makebox[\textwidth]{\rule{\textwidth}{0.4pt}}
\\[0.25em]
#2
\end{tcolorbox}
}

\begin{document}
\fancyhead{}

\input{Headings/head.tex}

\renewcommand{\shortauthors}{Li et al.}
\input{Headings/abstract.tex}

\settopmatter{printacmref=false}
\maketitle

\input{Sections/1.introduction.tex}
\input{Sections/2.Related_work}
\input{Sections/3.Method}
\input{Sections/4.Experemental_setting}
\input{Sections/5.Experimental_results}
\input{Sections/6.Conclusions}

\normalem
\balance
\bibliographystyle{ACM-Reference-Format}
\bibliography{aaai2026}
\clearpage

\input{Sections/8.appendix}
\end{document}

%% file: definitions.tex
\acrodef{PPO}{proximal policy optimization}
\acrodef{DPO}{direct preference optimization}
\acrodef{SFT}{supervised fine-tuning}
\acrodef{DAP}{direct alignment from preference}
\acrodef{SiLC}{sequence likelihood calibration}
\acrodef{IPO}{identity policy optimisation}
\acrodef{AI}{Artificial Intelligence}
\acrodef{LLM}{Large Language Model}
\acrodef{NDCG}{Normalized Discounted Cumulative Gain}
\acrodef{MAP}{Mean Average Precision}
\acrodef{MRR}{Mean Reciprocal Rank}

%% file: Headings/head.tex

%

\title{Are Recommenders Self-Aware? Label-Free Recommendation Performance Estimation via Model Uncertainty}

\author{Jiayu Li}
\authornote{These authors contributed equally to this work.}
\email{lijiayu997@gmail.com}
\affiliation{%
  \institution{DCST, Tsinghua University}
  \city{Beijing}
  \country{China}
}

\author{Ziyi Ye}
\authornotemark[1]
\email{yeziyi1998@gmail.com}
\affiliation{%
  \institution{DCST, Tsinghua University}
  \city{Beijing}
  \country{China}
}

\author{Guohao Jian}
\email{jgh22@mails.tsinghua.edu.cn}
\affiliation{%
  \institution{DCST, Tsinghua University}
  \city{Beijing}
  \country{China}
}

\author{Zhiqiang Guo}
\email{georgeguo.gzq.cn@gmail.com}
\affiliation{%
  \institution{DCST, Tsinghua University}
  \city{Beijing}
  \country{China}
}

\author{Weizhi Ma}
\email{mawz@tsinghua.edu.cn}
\affiliation{%
  \institution{AIR, Tsinghua University}
  \city{Beijing}
  \country{China}
}

\author{Qingyao Ai}
\email{aiqy@tsinghua.edu.cn}
\affiliation{%
  \institution{DCST, Tsinghua University}
  \city{Beijing}
  \country{China}
}

\author{Min Zhang}
\email{z-m@tsinghua.edu.cn}
\affiliation{%
  \institution{DCST, Tsinghua University}
  \city{Beijing}
  \country{China}
}

%% file: Headings/abstract.tex
\begin{abstract}
Can a recommendation model be self-aware? 
This paper investigates the recommender's self-awareness by quantifying its uncertainty, which provides a label-free estimation of its performance. 
Such self-assessment can enable more informed understanding and decision-making before the recommender engages with any users.
To this end, we propose an intuitive and effective method, probability-based \textbf{Li}st \textbf{D}istribution \textbf{u}ncertainty~(\textbf{LiDu}).
LiDu measures uncertainty by determining the probability that a recommender will generate a certain ranking list based on the prediction distributions of individual items.
We validate LiDu's ability to represent model self-awareness in two settings: (1) with a matrix factorization model on a synthetic dataset, and (2) with popular recommendation algorithms on real-world datasets.
Experimental results show that LiDu is more correlated with recommendation performance than a series of label-free performance estimators. 
Additionally, LiDu provides valuable insights into the dynamic inner states of models throughout training and inference. 
This work establishes an empirical connection between recommendation uncertainty and performance, framing it as a step towards more transparent and self-evaluating recommender systems.

\end{abstract}

%% file: Sections/1.introduction.tex
\section{Introduction}
\label{sec:intro}

%


\begin{figure}[t]
    \centering
    \includegraphics[width=\linewidth]{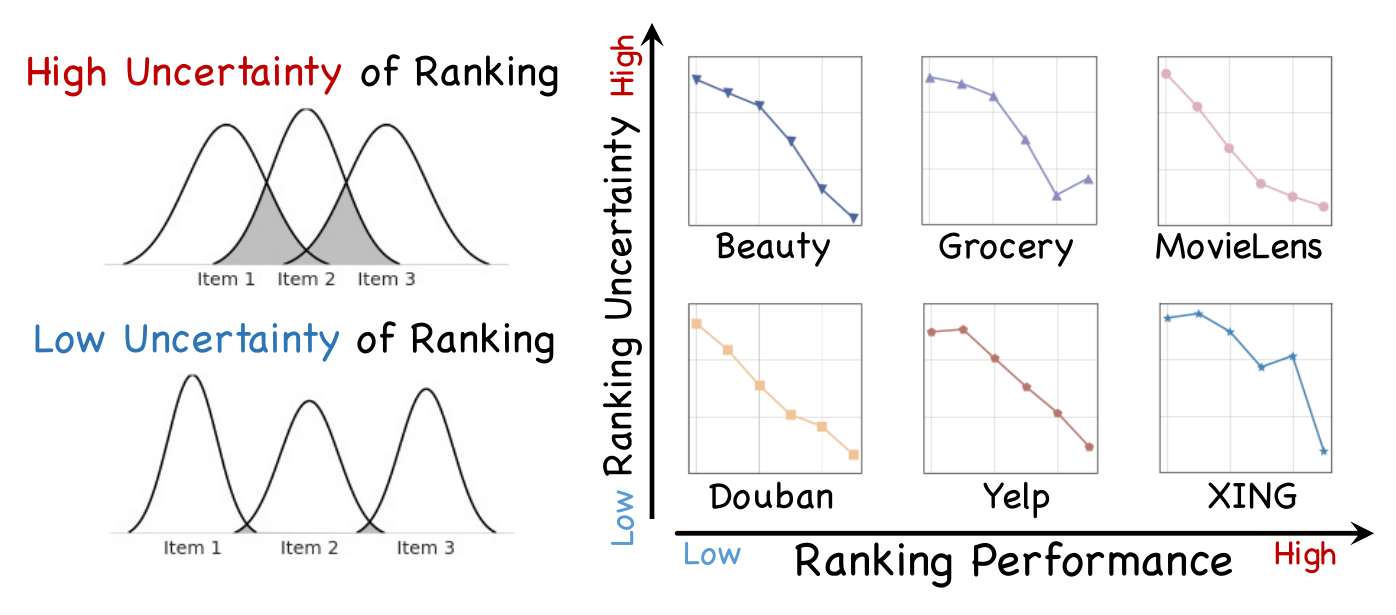}
    \setlength{\abovecaptionskip}{0pt}
    \setlength{\belowcaptionskip}{0pt}
    \caption{Empirical results suggest ranking performance of recommenders is negatively correlated with our proposed list distribution uncertainty~(LiDu) across various datasets.}
    \label{fig:illustration}
    \vspace{-15pt}
\end{figure}

Recommender systems (RecSys), as a widespread application of modern machine learning, utilize users' historical interactions to forecast user-preferred items.
Many of the modern RecSys operate as complex "black-box" models. 
This indicates that such models typically lack a crucial capability to distinguish between high-confidence recommendations and mere guesses, which could negatively impact user experience.
While existing efforts have focused on designing recommenders that perform well according to label-based evaluation metrics, a crucial question persists: are these models self-aware of their predictions?
In other words,  is it possible to quantify a recommender's predictive uncertainty for self-assessment?


Intuitively, lower uncertainty in RecSys's predictions indicates greater confidence in the user's preferences, which is likely to relate to better performance.
This hypothesis has been explored in scenarios such as Bayesian inference and image recognition~\cite{guo2017calibration}.
For example, \citet{guo2017calibration} empirically demonstrates that the performance of an image recognition model is negatively correlated with its uncertainty. 
They employ the term ``calibration'' to indicate a model that accurately measures confidence in its predictions.
With the advancement of modern neural networks, \citet{minderer2021revisiting} further finds that the latest image recognition models are both highly accurate and well-calibrated.

Despite the success in such scenarios, few studies have thoroughly investigated the correlation between uncertainty and performance in recommendation tasks.
As a label-free, model-focused measurement, uncertainty can be calculated prior to the user's interaction with the recommendation list.
Hence, it will be possible to anticipate the potential effects of recommendations in advance.
For example, it can be used to detect poor-quality recommendations, augment data, and assist in the ensemble of recommenders~\cite{jiangconvolutional2020,huang2025desocial}. 
Moreover, it may even help bridge the gap between offline and online evaluations. 
Therefore, it is crucial to investigate the correlation between the recommender's inner uncertainty and its actual performance measured by user feedback.

However, applying uncertainty quantification in the Top-N recommendation is non-trivial.
In classification tasks such as image recognition, uncertainty is measured by the variance of the prediction score using methods such as Bayesian Neural Network~\cite{blundell2015weight}, MC Dropout~\cite{gal2016dropout}, and Deep Ensembles~\cite{lakshminarayanan2017simple}.
However, Top-N recommendations aim to achieve better ranking rather than better score predictions, indicating that uncertainty should be considered at the \textbf{list} level rather than individual predictions.
Additionally, recommendation systems involve real-world, complex user features, which make it difficult to measure the uncertainty of the model due to factors such as model misspecification~\cite{masegosa2020learning} and data noise.
Existing studies for uncertainty in Top-N ranking tasks mainly focus on intra-list re-ranking with point-wise uncertainty~\cite{heusspredictive2023,yangmitigating2023,wangrethinking2023} rather than inter-list performance estimation.
Measuring uncertainty in Top-N recommendation tasks and understanding its role in evaluating RecSys remains an open problem.

To resolve this problem, we propose \textbf{Li}st \textbf{D}istribution \textbf{u}ncertainty (\textbf{LiDu}) to quantify uncertainty in Top-N recommendations, which measures the model's uncertainty in generating the ranking list considering both mean and variances of the predictions of Top-N items.
We first present a matrix factorization~(MF) experiment on synthetic recommendation scenarios to illustrate that LiDu is prone to be negatively correlated with model performance.
The MF experiment also shows that the negative correlation is strengthened with increased density and frequency of interaction data.
Motivated by the MF experiment, we conduct experiments on six real-world datasets with five commonly used recommendation models.
On all real-world datasets, {LiDu} exhibits a strong negative correlation with the performance, as shown in Figure~\ref{fig:illustration} using the recommender of BPRMF~\cite{rendle2012bpr}. 
Such correlation also demonstrates robust and superior performance compared to other potential estimators, such as training losses, statistical features, and uncertainty measured independently of the list.
Furthermore, the proposed {LiDu} outperforms other baselines for both active and inactive users, demonstrating the robustness of list-wise uncertainty as a performance estimator.
Finally, we reveal a series of properties of {LiDu} related to the training and inference processes, showing a pathway to understand the recommender's behavior beyond performance measurements.

In summary, our contributions are as follows:
\begin{itemize}[leftmargin=5pt]
    \item We propose \textbf{Li}st \textbf{D}istribution \textbf{u}ncertainty~(\textbf{LiDu}), to quantify uncertainty for the Top-K recommendation task.
    \item We illustrate a negative correlation between the proposed uncertainty and ranking models' performance through a synthetic MF experiment and present several observations regarding the density and frequency of interaction data.
    \item We test LiDu on six real-world recommendation datasets and five classical recommenders. 
    Results present that LiDu can be used as a robust performance estimator and exhibits beyond-accuracy properties related to recommender behaviors.
\end{itemize}

%% file: Sections/2.Related_work.tex
\section{Related Work}

\subsection{Uncertainty Quantification}
Uncertainty is typically categorized based on two factors: aleatoric uncertainty from the data noises, and epistemic uncertainty due to models' lack of knowledge~\cite{kendall2017uncertainties}.
To quantify such uncertainty, existing literature has proposed methods based on single deterministic~\cite{sensoy2018evidential,malinin2018predictive}, Bayesian~\cite{gal2016dropout}, ensemble~\cite{yang2020auto}, and test-time augmentation~\cite{hekler2023test}. 
As deep learning models become popular in recommendations, recent studies have paid attention to uncertainty quantification~(UQ) in deep RecSys~\cite{jiangconvolutional2020,zhouuncertaintyaware2023}.
For example, existing study have utilize UQ methods to intervene the output predictions to improve ranking performance~\cite{wangrethinking2023,cohennot2021} or debias the ranking resultss~\cite{heusspredictive2023,yangmitigating2023,huuncertainty2023}. 
However, such UQ is confined to point estimation rather than the uncertainties of ranking lists. 
While a few previous studies~\cite{coscratoestimating2023,coscrato2022recommendation} have intuitively adopted the average prediction variance of top-N items as a measure of uncertainty, this paper introduces a more direct, list-wise approach to quantifying uncertainty.
We show that such a metric is an effective performance estimator for top-N recommendation, thereby offering valuable insights for evaluating and improving recommendation strategy.

\subsection{Label-free Performance Estimation in IR}

Performance evaluations in Information Retrieval~(IR) have long been relying on labels such as users' implicit and explicit feedback~\cite{richmond1963review,lu2018between,ye2022don}.
For example, in Top-N ranking, performance is judged by the rank of positive items with metrics such as Hit Rate, \ac{NDCG}~\cite{jarvelin2002cumulated}, and \ac{MRR}~\cite{voorhees1999trec}.
However, obtaining these labels requires labor-intensive manual annotation or online user behavior collection that risks compromising user satisfaction~\cite{bauer2024exploring,ye2024relevance}. 
In the search scenario, a series of works have explored label-free estimation of search performance of queries, termed \textit{Query Performance Prediction}~(QPP) task~\cite{faggioli2023query}.
However, label-free performance estimators for RecSys still remains to be explored.
To address this gap, we investigate various measurements, including uncertainty, loss on the training set, and possible QPP baselines and show that uncertainty is both effective and robust.

%% file: Sections/3.Method.tex
\section{Methodology}
\label{sec:method}

\subsection{List Distribution Uncertainty} \label{sec:method_uncertainty}
Uncertainty quantifies the negative likelihood of certain outcomes when some aspects of the system are not exactly known.
Conventionally, uncertainty is calculated on a single prediction, i.e., $-P(y|x)$, where $x$ is the input and $y$ is the predicted score input $x$.
This approach acknowledges that scores or ratings assigned to items are often subject to variability due to factors such as data noise, model prediction errors, and inherent randomness in user preferences.
However, the target of Top-N recommendation is to generate an optimal ranking list of candidate items.
Therefore, the scores' absolute values are unimportant;
what matters is the comparative relationship between the scores.
Hence, we propose List Distribution Uncertainty~(\textbf{LiDu}) to calculate uncertainty for ranking as the negative likelihood of the most probable ranking generated by the model, i.e., $P(y_1>y_2>y_3>...>y_n|x)$. 

In this context, each $y_i$ is modeled as a Gaussian distribution characterized by a mean (the expected score) and a variance (the variability of the score). 
For any two items $ i $ and $ j $, the probability that item $ i $ ranks higher than item $ j $ is computed as follows:

\begin{equation}
P(r_i > r_j) := \pi_{i,j} = \int_{0}^{\infty} N(s \mid s_i - s_j, \sigma_i^2 + \sigma_j^2) \, ds
\end{equation}

where $N$ indicates a Gaussian distribution, $ s_i $ and $ s_j $ are means of the scores for items $ i $ and $ j $, $ \sigma_i^2 $ and $ \sigma_j^2 $ are their variances, which can be obtained via UQ methods on a single prediction.

The probability of generating a specific ranking for top $ K $ items is:

\begin{equation}
\small
P(r_1=1, r_2=2, \ldots, r_K=K) = \prod_{j>1} \pi_{1,j} \cdot \prod_{j>2} \pi_{2,j} \cdot \ldots \cdot \prod_{j>K} \pi_{K,j}
\label{eq:prob}
\end{equation}

This approach effectively combines pairwise ranking probabilities to derive the likelihood of a ranking list. 
Finally, the uncertainty for ranking, {LiDu}, is mathematically equivalent to the negative likelihood of the generated list.

\begin{equation}
    LiDu = -\log P(r_1=1, r_2=2, \ldots, r_K=K)
\end{equation}

Conventionally, recommendation models learn a score $s$ for a pair of items and a user without knowing the variance $\sigma$.
Therefore, we adopt UQ methods to measure the variance, including Monte Carlo~(MC) dropout~\cite{gal2016dropout}, deep ensembles~\cite{lakshminarayanan2017simple}, and variational Bayesian~\cite{harrisonvariational}.

\paragraph{MC dropout} 
MC dropout involves using dropout at the final layer of the model during inference to approximate Bayesian inference. 
Specifically, at inference time, the model runs $T$ forward passes with a dropout layer to obtain a set of predictions $\mathcal{S}=\{s_{1}, s_2, ..., s_T\}$.
And $\sigma$ is estimated by the variances of $\mathcal{S}$: $\sigma=1/T\cdot \sum_{t=1}^T(s_i-\mu)^2$, where $\mu$ is the mean of $\mathcal{S}$.
Since the dual-tower structure is commonly adopted in Top-N recommendations, and the ranking task involves a single user and multiple items, we only add a dropout layer to the user embedding representation to reduce computational overhead.
This indicates that the $T$ runs can share the same computation procedures and results, differing only in the user embedding layer.
For recommendation systems, especially those with complex networks, the overhead introduced is almost negligible.

\paragraph{Deep ensemble} Deep ensemble involves training multiple models independently and using their diversity to estimate uncertainty. 
For $T$ models trained with different initialization parameters, we obtain a set of predictions $\{s_{1}, s_2, ..., s_T\}$ and compute the variance $\sigma$ in the same way as MC dropout.
The difference is that the computational overhead for deep ensembles during both inference and training is significantly larger than that of MC dropout.

\paragraph{Variational Bayesian} 
Variational Bayesian is a single-pass Bayesian method for point-wise uncertainty quantification.
Specifically, it replaces the last layer of the recommender with a Bayesian weight matrix, which outputs both a score and the prediction variance.
Then, the recommender is trained using a marginal likelihood loss, which enables the output score and variance to adhere to their true distribution.

\begin{figure*}[t]
    \centering
    \begin{subfigure}{0.23\linewidth}\includegraphics[width=\linewidth]{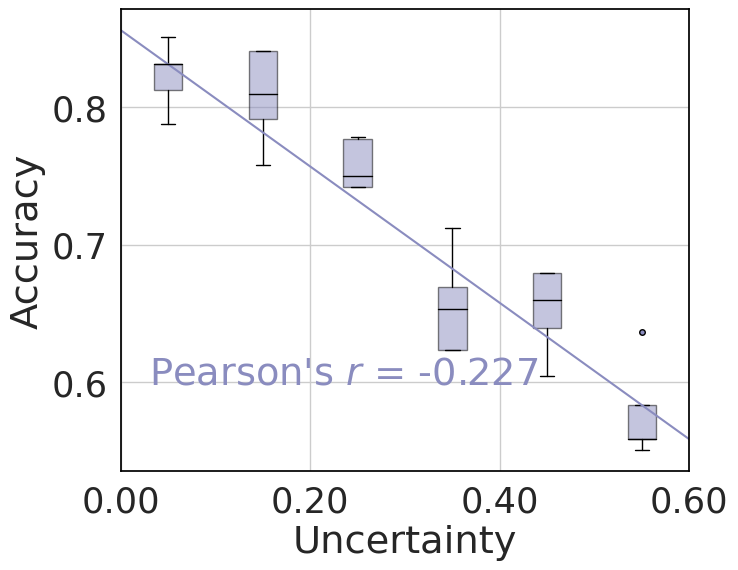} 
        \caption{Acc w.r.t Uncertainty\label{fig:four_a}}
    \end{subfigure}
    \begin{subfigure}{0.23\linewidth}\includegraphics[width=\linewidth]{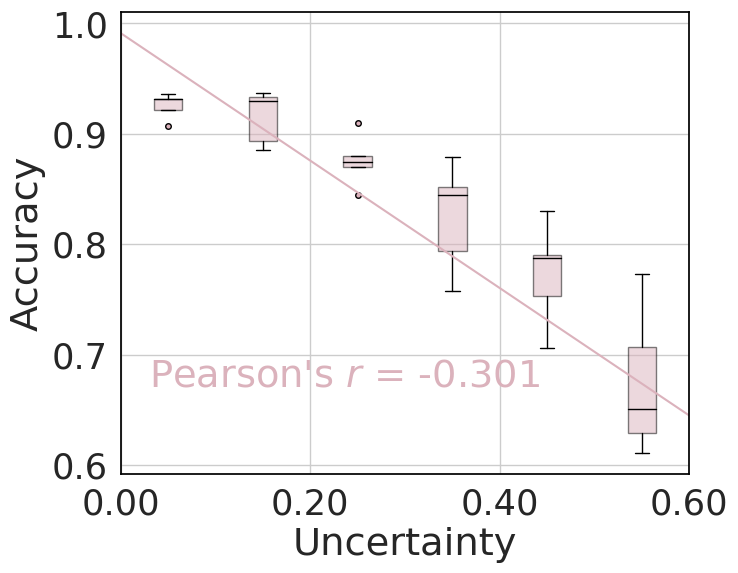} 
        \caption{Acc w.r.t Uncertainty\label{fig:four_b}}
    \end{subfigure}
    \begin{subfigure}{0.225\linewidth}\includegraphics[width=\linewidth]{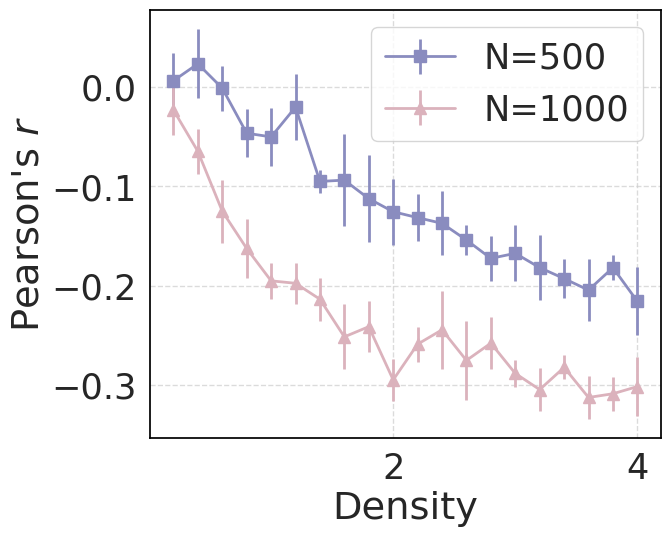} 
        \caption{Pearson's r w.r.t. Density\label{fig:four_c}}
    \end{subfigure}
    \begin{subfigure}{0.235\linewidth}\includegraphics[width=\linewidth]{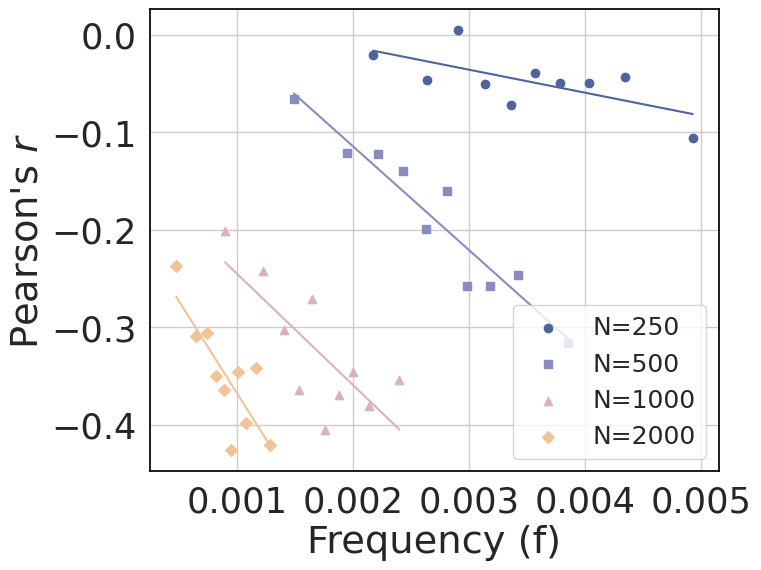} 
        \caption{Pearson's r w.r.t. Frequency\label{fig:four_d}}
    \end{subfigure}
    \vspace{-10pt}
    \caption{
    Results of the matrix factorization experiment on the synthetic data: (a)~Uncertainty presents a negative correlation with prediction accuracy~(for density = 4.0\% and N =500).
    (b)~For density = 4.0\% and N=1000.
    (c-d)~This Negative correlation enhances as the density and interaction density increase.
    }
    \label{fig:four_images}
\end{figure*}

\subsection{Theoretical Motivation}
\label{ssec:method_motivation}
In this section, we theoretically analyze whether our proposed uncertainty can serve as an indicator of prediction performance.
We begin by drawing inspiration from physics, using Boltzmann's theory to explain the motivation that uncertainty is negatively correlated with model performance. 
Next, we theoretically analyze the reasons for this negative correlation between uncertainty and model performance, based on a matrix factorization~(MF) task that resembles task scenarios of recommendation systems.

\subsubsection{Insights from the Boltzmann Theory}
Given an input $x$ and its corresponding output list $\pi=\{y_1, y_2, ..., y_n\}$, we define a quality function $Q(\pi)$ which quantify the effectiveness of the ranking list $\pi$.
$Q$ could be ranking-based metrics such as \ac{NDCG} and \ac{MRR}.
The learning process of the model $P(\pi|x,\theta)$ typically involves finding the state with the lower prediction error for $P(\pi^*|x,\theta)$, and usually corresponding to higher performance $Q(\pi|x,\theta)$, where $\pi^*$ is the optimal ranking list, $\theta$ is the model's parameter.
From the perspective of statistical physics, a prediction from a learning system with lower error corresponds to a state with lower energy:
\begin{equation}
Q(\pi|x,\theta) \propto -E(\pi) \label{formula:1}
\end{equation}
where $E$ is the energy for the model when output $\pi$.
Here we consider $\pi_i$ as one of the ranking distribution among all possible permutations and refer to the Boltzmann distribution~\cite{landau2013statistical}, which describes the probability of a system being in a state with energy $E$ at thermal equilibrium:
\begin{equation}
P(E(\pi_i)) = \frac{e^{-\beta E(\pi_i)}}{Z} \label{formula:2}
\end{equation}
where $E(\pi_i)$ is the energy of the state with ranking distribution $\pi_i$, $\beta = \frac{1}{k_B T}$ is the inverse temperature, with $k_B$ being the Boltzmann constant and $ T $ the temperature, $ Z $ is the partition function, defined as $Z = \sum_{j} e^{-\beta E(\pi_j)}$.
The exponential factor $e^{-\beta E(\pi_i)}$ indicates that lower energy states are more probable, and the partition function $Z$ normalizes the probabilities.
From Eq.~\ref{formula:1} and Eq.~\ref{formula:2}, we can make a hypothesis that for a sufficiently effective model, its prediction performance $Q(\pi|x,\theta)$ for input $x$ is negatively correlated with its prediction probabilities.
Motivated by this, we propose the Boltzmann hypothesis for ranking distribution uncertainty:

\hypbox{Boltzmann Hypothesis for Distribution Uncertainty\label{hypothesis}}{A sufficiently effective learning system will produce predictions whose distribution uncertainties are negatively correlated with their prediction accuracy.
}

Relevant evidence of the Boltzmann hypothesis for prediction uncertainty is the investigation on expected calibration error~(ECE).
For example, \citet{guo2017calibration} empirically shows that, in a typical classification task, as the model is trained to perform better, it also tends to be better calibrated. 
We extend this conclusion to ranking tasks and use uncertainty to measure the model's confidence in the ranking list. 
Below, we also present experiments on synthetic tasks related to recommendations to verify the above hypotheses.

\subsubsection{Analysis on an MF task}
We test the Boltzmann hypothesis for prediction uncertainty on a matrix factorization~(MF) task, which resembles a recommendation task.
The MF experiment aims to learn two separate embeddings $X=\{x_1, x_2,...,x_N\}$ and $Y=\{y_1,y_2,...,y_n\}$ such that the interaction $z_{i,j}$ between entities $i$ and $j$ can be approximated by a function $f(x_i, y_j)=x_i^T y_j$.
Here, $Z=\{z_{1,1},z_{1,2},...,z_{N,N}\} \in \mathbb{R}^{N\times N}$, $\{X,Y\}\in \mathbb{R}^{N\times d}$, where $d$ is the dimension of the embeddings.
This task resembles a recommendation task in which $X$ and $Y$ can represent user and item embeddings, respectively, and $Z$ represents their interaction.

Let $f(x_i, y_j)$ learn a predictive distribution for the continuous target value $z_{i,j}$, which we assume to be Gaussian: $p(z_{i,j}|x_i, y_j) = \mathcal{N}(\mu_{i,j}, \sigma^2_{i,j})$. The model's prediction is the mean $\hat{z}_{i,j} = \mu_{i,j}$, and its uncertainty is the variance, $U_{i,j} = \sigma^2_{i,j}$. 
Assume that the model is trained by minimizing the Negative Log-Likelihood (NLL) for each data point, where the loss $L_{i,j}$ is:
\begin{equation}
    L_{i,j} = \frac{(\hat{z}_{i,j} - z_{i,j})^2}{2\sigma^2_{i,j}} + \frac{1}{2}\log(2\pi\sigma^2_{i,j})
\end{equation}
To find the optimal variance for a given prediction error, we take the derivative of $L_{i,j}$ for $\sigma^2_{i,j}$ and set it to zero:
\begin{align}
    \frac{\partial L_{i,j}}{\partial (\sigma^2_{i,j})} &= -\frac{(\hat{z}_{i,j} - z_{i,j})^2}{2(\sigma^2_{i,j})^2} + \frac{1}{2\sigma^2_{i,j}} = 0 \nonumber \\
    \implies \frac{1}{2\sigma^2_{i,j}} &= \frac{(\hat{z}_{i,j} - z_{i,j})^2}{2(\sigma^2_{i,j})^2} \nonumber \\
    \implies \sigma^2_{i,j} &= (\hat{z}_{i,j} - z_{i,j})^2 \label{eq:optimal_variance}
\end{align}
Equation \ref{eq:optimal_variance} shows that optimizing the NLL incentivizes the network to learn a variance that matches the squared error of its own prediction. 
This yields a direct relationship between the model's uncertainty output and its prediction performance:
\begin{equation}
    U_{i,j} = \sigma^2_{i,j} \approx (\hat{z}_{i,j} - z_{i,j})^2
\end{equation}
Consequently, high uncertainty ($U_{i,j}$) corresponds to a low prediction performance, and vice versa.

\subsection{MF Experiment on a Synthetic Dataset}

We test LiDu on a matrix factorization~(MF) experiment that resembles recommendation tasks on a synthetic dataset.
To construct the dataset, we synthesized $\{X^*,Y^*\}\in \mathbb{R}^{N\times d}$ with $X^* = \cos(R_x) + \sin(R_x) $ and $Y^*=\sin(T_y) - \cos(T_y)$, where $R_x \in \mathbb{R}^{N\times D}$ and $R_y \in \mathbb{R}^{N\times D}$ are randomly initialized matrices.
We then compute $Z^*$ using $z^*_{i,j}=x_i^T y_j$ for each $z^*_{i,j} \in Z^*$.
The MF task learns to reconstruct $Z^*$ by learning from a subset $Z_{train} \in Z^*$. 
The randomly initialized $X$ and $Y$ are trained based on $Z_{train}$ using a mean square error~(MSE) loss function:

\begin{equation}
    \mathit{L}=\sum_{z_{i,j} \in Z_{train}}(z_{i,j}-x_i^T y_j)^2
\end{equation}

$X^*$ and $Y^*$ are feasible optimal solutions for this task, indicating that the synthetic dataset has at least one convergent solution.
Next, we evaluate the model's performance on the test set $Z_{test} \in Z^*$ in a simplified ranking scenario with only two targets, calculating its ranking accuracy:
\begin{equation}
\small
    \mathit{Accuracy}=\frac{1}{N'}\sum_{i=1}^{N'}\mathit{sgn}({z_{i_1,j_1}-z_{i_2,j_2}})\cdot \mathit{sgn}({x_{i_1}^T y_{j_1} - x_{i_2}^T y_{j_2}})
\end{equation}
where $\text{sgn}$ is the sign function and $N'$ is the size of the test set.
We select N from $\{250,500,1000,2000\}$ and train the model with the Adam optimizer with learning rate and batch size set as 0.01 and 32, respectively.
During training, an early stopping strategy is used when the model does not improve on the validation set within 10 epochs. 
All training processes converged through early stopping.
The training set $Z_{train}$ is selected according to varying density, i.e., proportional to $Z^*$, ranging from 0.2\% to 4.0\%.
The size of the test $N'$ is set as 1000, and a validation set of size 1000 is constructed. 
Considering Zipf's law in recommendation interactions, i.e., a large proportion of interactions occur with top items and top users, we set the sampling probability of $z_{i,j} \propto 1/((i + j + 1)^{\alpha})$ when sampling the data sets.
$\alpha$ is empirically set as 5 in the experiment.
To calculate {LiDu}, MC dropout is adopted with $p=0.2$ and the number of forward passes $T=20$ to obtain the variance of the prediction score.
All experiments are run on five different random seeds, and the average result is reported.

Experimental results are presented in Figure~\ref{fig:four_images}.
From Figure~\ref{fig:four_a} and Figure~\ref{fig:four_b}, we observe that uncertainty is negatively correlated with the accuracy performance with N=500 and N=1000, with Pearson's $r$=-0.227 and Pearson's $r$=-0.301, respectively.
The proposed {LiDu} is also compared with the standard point-wise UQ method, which measures uncertainty by treating variance as the metric, i.e., the sum of variance for all items in the ranking list.
We observed that the standard method yields merely no correlation with accuracy performance, with Pearson's $r$ values of 0.022 for $N=1000$ and 0.029 for $N=500$, respectively.
The above results demonstrate the effectiveness of our proposed {LiDu} and reveal that point-wise methods for quantifying uncertainty are not applicable to recommendation scenarios.

Furthermore, we investigate this negative correlation in terms of dataset properties.
As shown in Figure~\ref{fig:four_c}, we observe that Pearson's $r$ becomes more strongly negative as the density of the training set increases.
This suggests that with more training data, the relationship between uncertainty and performance becomes more pronounced. 
A possible explanation is that the model becomes better calibrated with more training data.
Our experiment also considers different sampling frequencies for different interactions to synthesize the data distribution of a recommendation system.
As shown in Figure~\ref{fig:four_d}, higher sampling frequency interactions also exhibit a stronger negative correlation.
This implies that uncertainty is a more effective performance estimator for high-frequency users and items.

In summary, we draw the following conclusions from the experiments on the synthetic dataset:
\begin{itemize}[leftmargin=15pt]
    \item The proposed list-wise uncertainty is negatively correlated with the ranking accuracy.
    \item Such correlation is enhanced as the size of the training data increases and the frequency of interactions rises.
\end{itemize}

\subsection{Implementation for Top-N Recommendation}

In the Top-N recommendation task, the ranking perturbations of items further down the list are less important.
Therefore, to prevent the items at the tail from dominating {LiDu} and reduce the computational cost, Eq~\ref{eq:prob} is modified for Top-N recommendation as
\begin{multline}
    P(r_1=1, r_2=2, \ldots, r_N=N) = \\ \prod_{j=s_1}^L\frac{\pi_{1,j}}{p_1}\cdot\prod_{j=s_2}^L\frac{\pi_{2,j}}{p_2}\ldots \prod_{j=s_N}^L\frac{\pi_{N,j}}{p_N}
    \label{eq:rec_prob}
\end{multline}

Where $N$ and $L$ are hyper-parameters to determine the range of items considered in uncertainty calculation; $s_n=high\_bit(n)+n$ is the step to ensure that the calculation only takes into account the differences between items that are sufficiently far; $p_n=log_2(i+1)$ is a position bias to assign lower weights for items further down the ranking list inspired by NDCG~\cite{jarvelin2002cumulated}.

%% file: Sections/4.Experemental_setting.tex
\section{Experimental Setups}
To examine the correlation between our proposed {LiDu} and ranking performance in real-world recommendations, we conducted empirical studies with various recommenders and datasets.

\subsection{Datasets and Recommenders}
Six datasets and five recommenders are selected for the experiments, resulting in 30 different settings.
The selected datasets vary in domains, platforms, and sparsity, including \textbf{Amazon}~\cite{hou2024bridging}, \textbf{Movielens-1M}\footnote{https://grouplens.org/datasets/movielens/1m/}, \textbf{Douban-Book}~\cite{zhu2019dtcdr}, \textbf{XING}~\cite{abel2017recsys}, and \textbf{Yelp}\footnote{https://www.kaggle.com/datasets/yelp-dataset}.
The datasets involved in our study include the following:
\begin{itemize}[leftmargin=10pt]
    \item \textbf{Amazon}~\cite{hou2024bridging}: Product review datasets from various categories on the Amazon platform. We choose two commonly-used subsets, Grocery and Gourmet Food~(short as \textit{Grocery}) and \textit{Beauty}.
    \item \textbf{Movielens-1M}\footnote{https://grouplens.org/datasets/movielens/1m/}: A stable benchmark dataset with 1 million movie ratings.
    \item \textbf{Douban-Book}~\cite{zhu2019dtcdr}: A book rating dataset from \textit{Douban}, a large website for book and music sharing.
    \item \textbf{XING}~\cite{abel2017recsys}: A job recommendation dataset from RecSys Challenge 2017. We use it to recommend jobs for users based on user history.
    \item \textbf{Yelp}\footnote{https://www.kaggle.com/datasets/yelp-dataset}: A rating and review dataset for check-in interactions.
\end{itemize}

\begin{table}[]\footnotesize
    \centering
    \setlength{\abovecaptionskip}{0pt}
    \setlength{\belowcaptionskip}{2pt}
    \caption{Dataset Statistics. \# indicates \textit{The number of}.}
    \label{tab:dataset}
\begin{tabular}{lcccc}
\toprule
\textbf{Dataset} & \#\textbf{User} & \#\textbf{Item} & \#\textbf{Inter} & \textbf{Density} \\
\midrule
Amazon Beauty & 18,350 & 10,715 & 157,946 & 0.08\% \\
Amazon Grocery & 12,922 & 7,919 & 128,621 & 0.13\% \\
Movielens-1M & 6,039 & 3,384 & 998,718 & 4.89\% \\
Douban Book & 1,858 & 8,044 & 96,446 & 0.65\% \\
XING & 82,180 & 12,117 & 810,687 & 0.08\% \\
Yelp & 55,616 & 34,945 & 1,506,777 & 0.07\% \\
\bottomrule
\end{tabular} 
\vspace{-10pt}
\end{table}

For XING, we retained interactions where $\text{interaction\_type} > 0$.
The other datasets are all based on five-level ratings, so interactions with $\text{rating} \geq 3$ are retained. 
To ensure the recommenders are adequately trained, 5-core filtering is conducted on all datasets following previous works~\cite{wang2022target,hou2024bridging,kang2018self}. 
The statistics of processed datasets are shown in Table~\ref{tab:dataset}.

Since we focus on Top-N recommendations, dual-tower recommenders are selected, including both collaborative filtering~(CF) and sequential ones.
The selected recommenders are as follows:
\begin{itemize}[leftmargin=10pt]
    \item \textbf{BPRMF}~\cite{rendle2012bpr}: A simple matrix factorization CF method with pair-wise BPR loss.
    \item \textbf{LightGCN}~\cite{he2020lightgcn}: A CF recommender with a simplified graph convolutional network over user-item interactions.
    \item \textbf{SimpleX}~\cite{mao2021simplex}: A CF recommender that generates user profiles based on user IDs and historical interaction pooling, which optimizes through cosine contrastive loss (CCL).
    \item \textbf{SASRec}~\cite{kang2018self}: A sequential recommender using a self-attention module to balance long-term and short-term prediction.
    \item \textbf{TiMiRec}~\cite{wang2022target}: A sequential recommender to model users' multi-interest profiles and generate recommendations by distilled interest for the target item.  
\end{itemize}
As we aim to test the applicability of our proposed uncertainty to Top-K recommendation rather than compare SOTA recommenders, typical models are selected to include different information (collaborative filtering and sequential), architectures (Graph, Pooling, Attention, Multi-sequence, etc.), and training loss functions.

\begin{table}[]
    \centering
    \setlength{\abovecaptionskip}{0pt}
    \setlength{\belowcaptionskip}{2pt}
    \caption{The recommendation performances~(in terms of NDCG@1000) of all models over all datasets.}
    \label{tab:rec_performance}
\resizebox{\columnwidth}{!}{
    \begin{tabular}{l|ccccc}
    \toprule
\multicolumn{1}{l|}{\multirow{2}{*}{\textbf{Dataset}}} & \multicolumn{5}{c}{\textbf{Model}} \\
 & BPRMF & LightGCN & SimpleX & SASRec & TiMiRec \\
 \midrule
Grocery & 0.0798 & 0.0958 & 0.0873 & 0.1069 & 0.1066 \\
Beauty & 0.0774 & 0.0914 & 0.0856 & 0.0952 & 0.0965 \\
ML1M & 0.1650 & 0.1643 & 0.1456 & 0.2492 & 0.2559 \\
Douban & 0.0660 & 0.0617 & 0.0756 & 0.0862 & 0.0965 \\
XING & 0.3531 & 0.3572 & 0.3847 & 0.4305 & 0.3994 \\
Yelp & 0.0936 & 0.0952 & 0.0914 & 0.0982 & 0.1000 \\
\bottomrule
\end{tabular}}
\vspace{-10pt}
\end{table}

\subsection{Recommender Training}
Since both CF and sequential recommenders are considered, we adopt the \textit{leave-one-out} dataset splitting strategy, which is commonly used in sequential recommendation~\cite{wang2022target,kang2018self}: The last interaction of each user belongs to the test set, the second last for the validation set, and the remaining for the training set. 
All recommenders are implemented with the ReChorus toolkit~\cite{li2024rechorus2}. 
Most of the hyper-parameters follow the default parameters of the toolkit, and detailed hyper-parameter settings are all publicly available in our repository.
Adam optimizer is used for training, and BPR loss is adopted for all recommenders except SimpleX.
Evaluations are conducted over all items, and early stopping is adopted until NDCG@1000 on the validation set does not increase for 10 epochs.
The performances of all recommenders and datasets are shown in Table~\ref{tab:rec_performance}.
Overall, the sequential recommenders outperform CF models, which aligns with general experience. 
The following experiments and analyses are all based on these trained recommenders.

\subsection{Performance Estimation Setups}
We compare \textbf{LiDu} with other possible label-free estimators for ranking performances, including training loss and query performance prediction~(QPP) baselines.
The leave-one-out setting is adopted for the recommender's performance estimation, consistent with the strategy used during its training. 

\subsubsection{Baseline Methods}
\begin{itemize}[leftmargin=5pt]
    \item \textbf{loss}: Average loss of interactions for a user in the training set when the model converges.
    \item \textbf{SMV}~\cite{tao2014query}: A QPP method that predicts performance based on the deviation and magnitude of scores of Top-N items.
    \item \textbf{NQC}~\cite{shtok2012predicting}: A QPP method that uses the deviation of scores of Top-N items as an estimator of performance.
    \item \textbf{W-Graph}~\cite{arabzadeh2021query}: A QPP method that builds a pruned graph based on the similarity of Top items. All four weighted properties of the graph, WACC, WADC, WAND, and WD, are implemented and we reported the best-performing method as W-Graph.
\end{itemize}

In addition, we have also compared the LiDu approach with traditional methods of uncertainty quantification that focus on single predictions rather than distributions. 
We found that traditional point-wise uncertainty measurements exhibit insignificant correlation with the recommendation performance, both in the MF experiments and experiments on real-world recommendations. 
Consequently, we have chosen not to include these methods as a baseline.

\subsubsection{Evaluation Metrics}

NDCG@K~\cite{jarvelin2002cumulated} is used as the performance ground truth, where K=100 for XING and 1000 for others.
Evaluations are based on how well estimators are consistent with NDCG@K when comparing all ranking samples in the test set with the same model and dataset.
The following metrics are adopted:
\begin{itemize}[leftmargin=10pt]
    \item \textbf{Win Rate-$\delta$}: Given any two ranking lists $l_i$ and $l_j$, the probability that judgment by NDCG@K and the estimator are the same: $P(N@K_i<N@K_j|e_i>e_j+\delta),\quad \forall i,j$, where $e_1$ is the estimation value. $\delta$ is a pre-defined threshold to eliminate comparisons between close estimations. $\delta$ is set to 5\% of the test set size.
    The chance performance of Win Rate-$\delta$ is 0.5.
    \item \textbf{Pearson's r}: Pearson's r correlation between the estimators and NDCG@K on all test ranking samples.
    \item \textbf{sARE}~\cite{faggioli2023query}: Rank all test samples with NDCG@K and estimators respectively, and calculate the average differences of two ranks: $\sum_{i=1}^{n}|R_{N@K,i}-R_{e,i}|/n$.
    $n$ is the size of test data, and $R_{N@K}$ and $R_e$ are the ranking of NDCG@K and estimations in all samples. Due to page limitations, sARE results are presented in the supplementary information. 
\end{itemize}

\subsubsection{Implementation of Uncertainty}
MC Dropout, Ensemble, and variational bayesian are utilized to obtain the variances of prediction scores, denoted as \textbf{LiDu-dp}, \textbf{LiDu-en}, and \textbf{LiDu-vb}, respectively.
For LiDu-dp, since all recommenders follow dual-tower structures and the uncertainties reflect user understandings, a dropout layer is added to the user representations before the final cross layer between user and item.
Dropout probability is set to 0.2 with $T=50$ times forward passes.
For LiDu-en, five models with different random initializations are used. 
When generating uncertainty, we set $N=100$ and $L=1000$ in Eq.\ref{eq:rec_prob} for all datasets except XING. 
XING shows significantly higher performance, so $N=10$ and $L=100$ are selected.
LiDu-vb is implemented by replacing the last layer of recommenders with a Bayesian weight matrix.

\subsubsection{Implementation of QPP Methods}
For fair comparisons, QPP baselines are selected based on two principles: (1) Not relying on language models due to the lack of text in RecSys; (2) No extra neural models are needed.
Thus, statistics-based post-retrieval QPP models are selected as baselines, including SMV, NQC, and W-Graph.
We tune the number of top items for all baselines in $\{10,100,1000\}$, and find N=10 for XING and N=100 for all other datasets leads to the best estimation.
Moreover, the opposite numbers of all QPP are adopted to ensure the same trend as uncertainty, i.e., negatively correlated with performance.


%% file: Sections/5.Experimental_results.tex
\section{Experimental Results}
\label{sec:performance}

\subsection{Overall Performance}

\begin{table*}[]
    \setlength{\abovecaptionskip}{0pt}
    \setlength{\belowcaptionskip}{2pt}
    \caption{Overall performance estimation results on six datasets with five recommenders. WG is short for W-Graph. NDCG@K is used as the performance label. The negative Pearson's $r$ is reported to avoid repeated negative signs. $\uparrow$~($\downarrow$) indicates larger~(smaller) metric is better. The best performance is shown in \textbf{Bold}, and the second best performance is \underline{underlined}. $*,\dagger,\ddagger$ indicate significantly underperform LiDu-dp, LiDu-en, and LiDu-vb, respectively. }
    \label{tab:overall_result}
    \centering
\resizebox{0.9\textwidth}{!}{
    \begin{tabular}{ll||llll|lll||llll|lll}
\hline
\multirow{3}{*}{\textbf{Dataset}} & \multirow{3}{*}{\textbf{Method}} & \multicolumn{7}{c||}{\textbf{Win Rate-$\delta$ ($\uparrow$)}} & \multicolumn{7}{c}{\textbf{$-$Pearson's $r$ ($\uparrow$)}}  \\
& & \multicolumn{4}{c|}{\textbf{baseline}} & \multicolumn{3}{c||}{\textbf{LiDu}} & \multicolumn{4}{c}{\textbf{baseline}} & \multicolumn{3}{|c}{\textbf{LiDu}} \\
 &  & \multicolumn{1}{c}{loss} & \multicolumn{1}{c}{SMV} & \multicolumn{1}{c}{NQC} & \multicolumn{1}{c}{WG} & \multicolumn{1}{|c}{dp} & \multicolumn{1}{c}{en} & \multicolumn{1}{c||}{vb} & \multicolumn{1}{c}{loss} & \multicolumn{1}{c}{SMV} & \multicolumn{1}{c}{NQC} & \multicolumn{1}{c}{WG}& \multicolumn{1}{|c}{dp} & \multicolumn{1}{c}{en} & \multicolumn{1}{c}{vb} \\
\toprule
\multirow{4}{*}{Grocery} &  BPRMF &  0.659 &  0.634 &  0.724 &  0.538 &  \underline{0.727} &  0.702 &  \textbf{0.760} &  0.146 &  0.092 &  {0.252} &  0.140 &  \underline{0.261} &  0.218 &  \textbf{0.271} \\
 &  LightGCN &  0.619 &  0.605 &  0.656 &  0.661 &  \textbf{0.684} &  \underline{0.672} & 0.668 &  0.090 &  0.066 &  0.181 &  {0.201} &  \underline{0.212} &  0.166 & \textbf{0.235} \\
 &  SimpleX &  0.628 &  0.597 &  0.601 &  0.648 &  \underline{0.686} &  {0.679} &  \textbf{0.697} &  0.164 &  \textbf{0.236} &  0.200 &  0.129 &  {0.202} &  0.177 &  \underline{0.211} \\
 &  SASRec &  0.634 &  0.715 &  0.701 &  {0.722} &  \underline{0.728} &  0.719 &  \textbf{0.769} &  0.048 &  0.274 &  0.248 &  0.296&  \underline{0.311} &  {0.305}&  \textbf{0.354} \\
 &  TiMiRec &  0.614 &  0.709 &  0.683 &  \textbf{0.719} &  \underline{0.716} &  0.709 &  0.712 &  0.024 &  0.278 &  0.238 &  0.301 &  \underline{0.301} &  \textbf{0.303} &  0.300  \\
\midrule               
\multirow{4}{*}{Beauty} &  BPRMF &  0.669 &  0.637 &  0.691 &  0.709 &  \textbf{0.748} &  {0.731} &  \underline{0.732} &  0.139 &  0.090 &  {0.288} &  0.202 &  \underline{0.306} &  0.268 &  \textbf{0.321}  \\
 &  LightGCN &  0.542 &  0.510 &  0.635 &  0.703 &  \textbf{0.717} &  \underline{0.716} & 0.664 &  -0.007 &  0.116 &  0.215 &  \underline{0.278} &  \textbf{0.285} &  0.268 & 0.185 \\
 &  SimpleX &  0.643 &  0.633 &  0.635 &  {0.678} &  0.662 &  \underline{0.692} &  \textbf{0.722} &  {0.241} &  \textbf{0.276} &  0.241 &  0.175 &  0.168 &  0.200 &  \underline{0.270} \\
 &  SASRec &  0.646 &  0.693 &  0.668 &  \underline{0.722} &  0.715 &  {0.716} &  \textbf{0.728} &  0.074 &  0.200 &  0.233 &  \underline{0.270} &  0.241 &  {0.249} &  \textbf{0.339} \\
 &  TiMiRec &  0.632 &  0.697 &  0.658 &  0.712 &  \textbf{0.717} &  \underline{0.714} &  0.707 &  0.067 &  0.208 &  0.254 &  0.246 &  \underline{0.258} &  \textbf{0.285} &  0.242  \\
\midrule               
\multirow{4}{*}{ML-1M} &  BPRMF &  {\bfseries0.621} &  0.601 &  0.579 &  \underline{0.614} &  0.594 &  0.602 &  0.597 &  \underline{0.117} &  0.078 &  0.041 &  \textbf{0.118} &  0.102 &  0.090 &  0.089 \\
 &  LightGCN &  \textbf{0.636} &  0.595 &  0.611 &  0.589 &  {0.612} &  0.601 & 0.586 &  \textbf{0.165} &  0.030 &  0.113 &  0.063 &  \underline{0.146} &  0.107 &  0.027 \\
 &  SimpleX &  \textbf{0.636} &  0.584 &  0.582 &  \underline{0.617} &  0.593 &  0.569 &  0.584 &  \textbf{0.154} &  0.051 &  0.048 &  {0.135} &  0.112 &  0.033 &   \underline{0.144}\\
 &  SASRec &  0.596 &  0.569 &  \underline{0.644} &  0.568 &  0.599 &  {0.620} &  \textbf{0.652} &  0.091 &  0.006 &  \textbf{0.207} &  0.047 &  0.110 &  {0.155} &  \underline{0.192}  \\
 &  TiMiRec &  0.596 &  0.603 &  \textbf{0.663} &  0.569 &  0.590 &  {0.623} & \underline{0.650} &  0.073 &  0.110 &  \textbf{0.259} &  0.050 &  0.104 &  {0.157} &  \underline{0.186} \\
\midrule               
\multirow{4}{*}{\begin{tabular}[c]{@{}l@{}}Douban \\ Book\end{tabular}} &  BPRMF &  \underline{0.605} &  0.566 &  0.579 &  0.581 &   \textbf{0.610} &  0.564 &  0.570 &  0.116 &  0.083 &  0.065 &  0.087 &  \textbf{0.145} &  \underline{0.144} &  0.142 \\
 &  LightGCN &  \textbf{0.607} &  0.564 &  0.522 &  0.562 &  \underline{0.577} &  0.554 & 0.541 &  \underline{0.100} &  0.052 &  -0.059 &  {0.000} &  0.093 &  \textbf{0.113} & 0.000 \\
 &  SimpleX &  0.608 &  0.614 &  \underline{0.628} &  0.611 &  \textbf{0.630} &  0.615 &  0.619 &  0.112 &  0.153 &  0.156 &  0.118 &  {0.208} &  \textbf{0.223} &  \underline{0.213} \\
 &  SASRec &  0.571 &  0.608 &  {0.609} &  0.607 &  \underline{0.615} &  0.599 &  \textbf{0.631} &  0.032 &  0.104 &  0.151 &  0.083 &  \textbf{0.176} &  \underline{0.164} &  0.113  \\
 &  TiMiRec &  0.565 &  0.587 &  0.576 &  0.586 &  0.580 &  \textbf{0.614} &  \underline{0.599} &  0.058 &  0.087 &  0.088 &  0.055 &  {0.113} &  \textbf{0.248} &  \underline{0.129}  \\
 \midrule               
 \multirow{5}{*}{XING} &  BPRMF &  0.646 &  \textbf{0.660} &  0.608 &  0.469 &  0.642 &  0.602 &  \underline{0.649} &  0.219 &  \underline{0.254} &  0.117 &  0.000 &  \textbf{0.261} &  0.150 & 0.237 \\
 &  LightGCN &  \underline{0.612} &  0.480 &  0.587 &  0.581 &  \textbf{0.616} &  0.528 & 0.542 &  0.050 &  0.042 &   0.069 &  0.000 &  \textbf{0.163} &  {0.064} & \underline{0.132} \\
 &  SimpleX &  0.649 &  0.552 &  0.604 &  0.611 &  {0.616} &  \underline{0.683} & \textbf{0.859} &  \underline{0.335} &  0.054 &  0.116 &  0.140 &  0.164 &  0.330 &  \textbf{0.565}\\
 &  SASRec &  0.669 &  0.534 &  0.663 &  {0.751} &  \underline{0.786} &  0.542 & \textbf{0.860} &  0.120 &  0.041 &  0.247 &  {0.408} &  \underline{0.541} &  0.063 & \textbf{0.707} \\
 &  TiMiRec &  0.653 &  0.681 &  {0.690} &  0.627 &  \underline{0.724} &  0.594 & \textbf{0.858} &  0.193 &  0.252 &  {0.318} &  0.242 &  \underline{0.464} &  0.123 & \textbf{0.764} \\
 \midrule               
 \multirow{5}{*}{Yelp} &  BPRMF &  0.608 &  0.621 &  0.635 &  0.626 &  0.655 &  \textbf{0.659} & \underline{0.658} &  0.034 &  0.067 &  0.125 &  0.093 &  \textbf{0.139} &  \underline{0.127} & 0.107 \\
 &  LightGCN &  0.612 &  0.575 &  \underline{0.623} &  0.618 &  \textbf{0.654} &  0.537 & 0.585 &  0.049 &  0.066 &  \underline{0.109} &  0.092 &  \textbf{0.136} &  0.104 & 0.060 \\
 &  SimpleX &  \textbf{0.654} &  0.598 &  0.499 &  \underline{0.610} &  0.521 &  0.558 & 0.588 &  \textbf{0.134} &  0.053 &  -0.053 &  \underline{0.066} &  -0.049 &  0.010 & 0.029 \\
 &  SASRec &  0.615 &  0.598 &  \underline{0.658} &  0.651 &  0.652 &  \textbf{0.667} & 0.616 &  0.059 &  0.035 &  \underline{0.158} &  0.122 &  0.133 &  \textbf{0.158} & 0.073 \\
 &  TiMiRec &  0.594 &  0.626 &  \textbf{0.661} &  0.642 &  0.655 &  \underline{0.658} & 0.548 &  0.031 &  0.059 &  \textbf{0.166} &  0.138 &  \underline{0.148} &  0.143 & 0.011 \\
 \midrule
 \multicolumn{2}{c||}{Average} & $0.621^{*,\dagger,\ddagger}$ &	$\text{0.608}^{*,\dagger,\ddagger}$ &	$\text{0.629}^{*,\ddagger}$ &	$\text{0.630}^{*,\ddagger}$ &	\underline{0.654} &	0.635 &	\textbf{0.665} &	$\text{0.107}^{*,\dagger,\ddagger}$ & $\text{0.117}^{*,\dagger,\ddagger}$ &	$\text{0.159}^{*,\ddagger}$ &	$\text{0.143}^{*,\ddagger} $ &	\underline{0.198} &	0.172 &	\textbf{0.222} \\ 
\bottomrule
\end{tabular}
}
\vspace{-10pt}
\end{table*}

\begin{figure}[h]
    \centering
\includegraphics[width=0.95\linewidth]{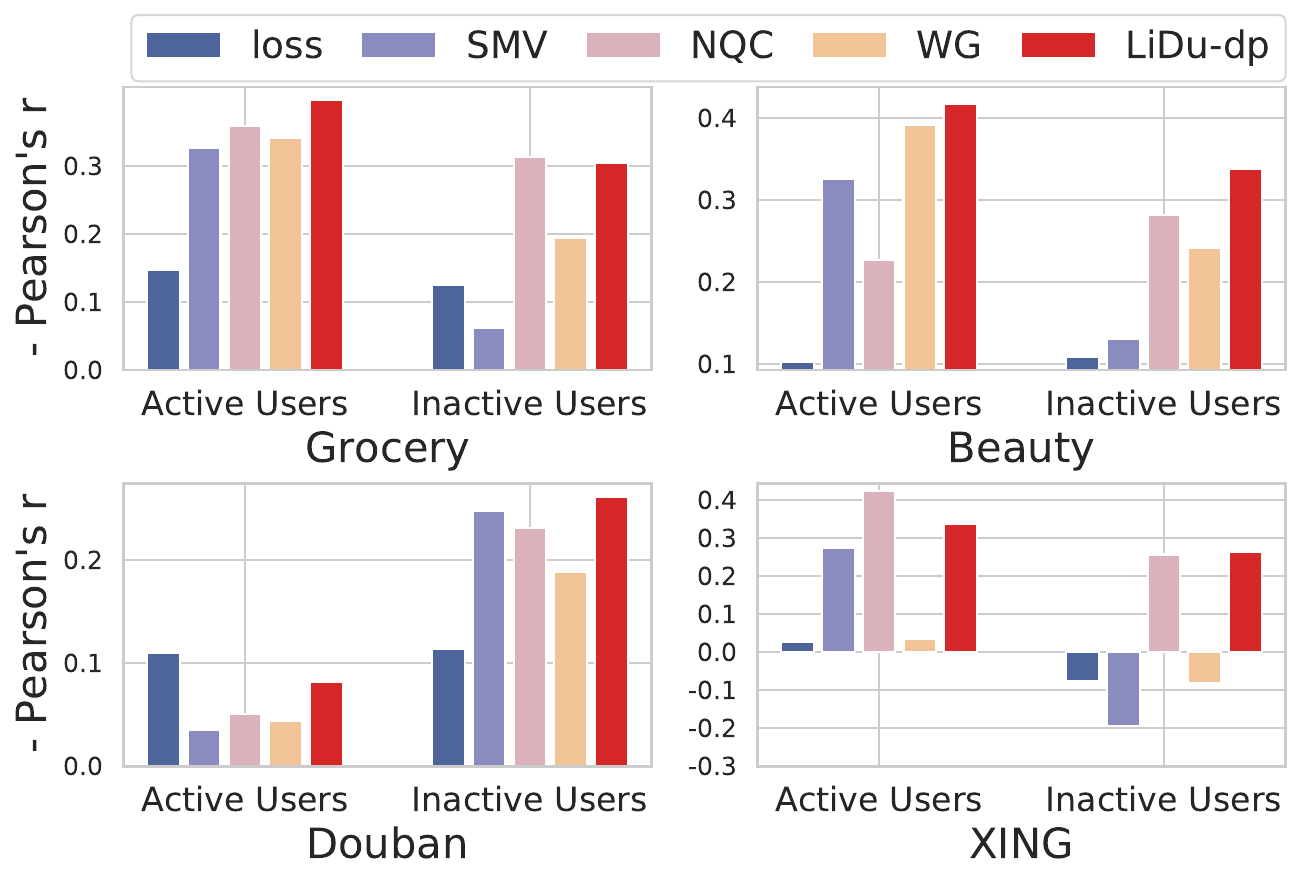}
    \caption{The absolute number of negative Pearson's $r$ between NDCG@K and estimations on active and inactive users on different datasets with BPRMF. Inactive user means a user with less than five interactions.}
    \label{fig:exp_activeness}
\end{figure}

The overall performance estimation results for \textbf{LiDU}, with different implementations including MC Dropout~(LiDU-dp), Deep Ensemble~(LiDU-en), and Variational Bayesian~(LiDU-vb), are presented in Table~\ref{tab:overall_result}.
In general, \textbf{LiDu} is more significantly correlated with the true performance than the baselines, including the training loss and a series of QPP baselines.
Especially, LiDu-vb and LiDu-dp have the best and second-best performance, respectively (with Pearson's $r$ of 0.222 and 0.198, and Win Rate-$\delta$ of 0.665 and 0.654).
The best performance is achieved by either LiDu-dp, LiDu-en, or LiDu-vb in 21 out of 30 experiments, with at least the second-best performance in 26 experiments regarding win rate.
Among all implementations of LiDU, LiDU-en performs the worst due to its instability with different training checkpoints. In contrast, LiDU-vb and LiDU-dp utilize the same checkpoint but measure output uncertainty through internal mechanisms.
On the other hand, none of the baselines exhibits consistent good performance across different methods and datasets.
Among them, W-Graph exhibits a similar tendency to LiDu, as it also reflects the uncertainty of recommenders. 
However, it generally underperforms LiDu because it calculates uncertainty based solely on the predicted score.

When comparing the effectiveness of label-free performance estimation across different datasets and recommenders, the highest win rates can exceed 0.6 and even surpass 0.7 on the Grocery, Beauty, and XING datasets. 
This indicates that Top-N recommendation performance is predictable in general, even though the performance of individual data points may be uncertain.
The performance predictability varies with different datasets, while for the same dataset, the best estimation effectiveness of different recommenders does not vary much.
This indicates that the performance of a sufficiently effective learned RecSys on one predictable dataset can be estimated by uncertainty, which is consistent with our MF experiments on the synthetic dataset.

Meanwhile, LiDu does not perform well in some cases, such as on the ML-1M dataset and using SimpleX on the Yelp dataset.
ML-1M is much denser than other datasets and involves a smaller gap between the training and validation sets. 
Consequently, the model can learn more information from the training set, making \textit{loss} a reliable estimator.
When we down-sample Ml-1M to 1/50 density, the effectiveness of all estimators decreases, while LiDu outperforms all baselines, including loss.
LiDu's poor performance for SimpleX on Yelp may be due to the large number of items in the Yelp dataset.  
SimpleX uses the cosine contrastive loss~(CCL) for optimization, which optimizes less for hard samples, leading to more inaccurate predictions for items with lower performance. 
Therefore, uncertainty, SMV, and NQC, which only use the scores of top items, all perform worse. 

\begin{figure*}[t]
    \centering
    \includegraphics[width=\textwidth]{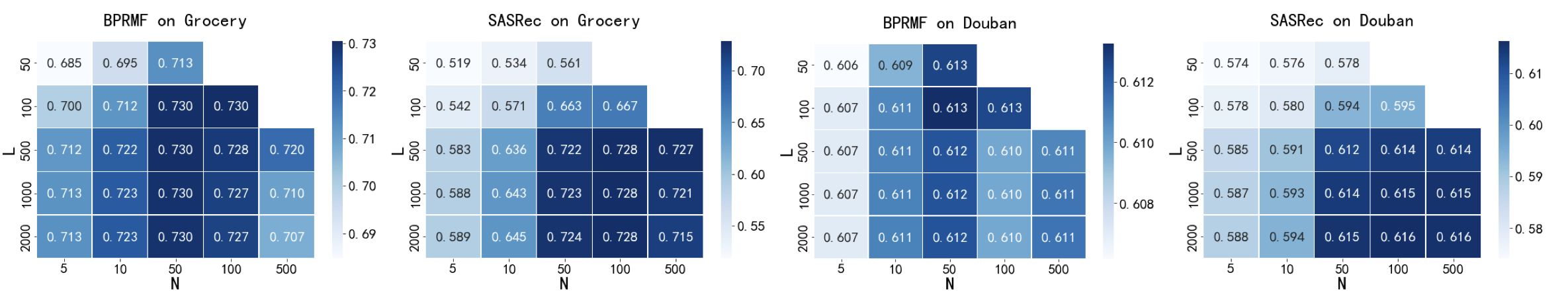} 
    \caption{The win rate of LiDu-dp estimating NDCG@1000 under different values of $N$ and $L$.}
    \label{fig:performance}
\end{figure*}

\subsection{Sensitive analysis}
As a performance estimator, LiDu doesn’t contain many hyperparameters, mainly $N$ (the number of top items for calculating probability) and $L$ (the maximum number of items considered for pairwise probability for each of $N$ items). 
The estimation accuracy (in terms of win rate of LiDu-dp estimating NDCG@1000) changing with $N$ and $L$ is shown in Figure \ref{fig:performance}. 
For simplicity, we only present the results on the two most classical methods and two datasets. 
The results show that when $N$ and $L$ are too small, the win rate is low. 
However, when $L$ reaches hundreds, the win rate becomes stable, which is because we introduce step and position bias (Equation \ref{eq:rec_prob}) to ensure the stability of uncertainty. 
Therefore, we use fixed hyperparameters~($N$=10 for XING and $N$=100 for all other datasets, $L$=1,000 for all datasets) in the main experiments without tuning.

\subsection{Further Analysis}

The experiment on the synthetic dataset~(see Figure~\ref{fig:four_d}) suggests that interaction frequencies are positively related to the performance estimation effectiveness of uncertainty. 
In this analysis, we explore whether there is a similar tendency in the real-world datasets.
We selected LiDU-dp for this analysis because it uses the same checkpoint as the baselines to fairly quantify uncertainty.

As uncertainties are analyzed from the users' perspective, we focus on the frequencies of users' interactions, i.e., users' activeness.
Users are grouped into active and inactive ones according to whether they have more than 5 interactions in the training set. 
The correlations between each estimator and NDCG@K for these two groups of users are shown in Figure~\ref{fig:exp_activeness}.
Results on Yelp and ML-1M are not presented because the proportions of inactive users in these datasets are less than 1\%.
In all datasets except Douban, the correlation between performance and uncertainty is higher in the active users group.
Moreover, LiDu-dp outperforms baselines, especially on inactive users, showing its potential for evaluating cold users.
The abnormal phenomenon observed on Douban can be attributed to its book recommendation scenario, where active users can display more dynamic interests than inactive users.
Specifically, the average interest diversity $d_u$ for active users is 0.738 and 0.650 for inactive users in Douban, which trend is different from other datasets~(see Eq.~\ref{eq:interest_dynamic} for the definition of $d_u$).
Therefore, it is more challenging to estimate the performance of active users, resulting in poor performance estimation results. 

\subsubsection{User Activeness and Estimation Accuracy}
\label{sec:User Activeness and Estimation Accuracy}

\subsubsection{Beyond-Accuracy Properties of Uncertainty}

\begin{figure}
    \centering
    \includegraphics[width=0.95\linewidth]{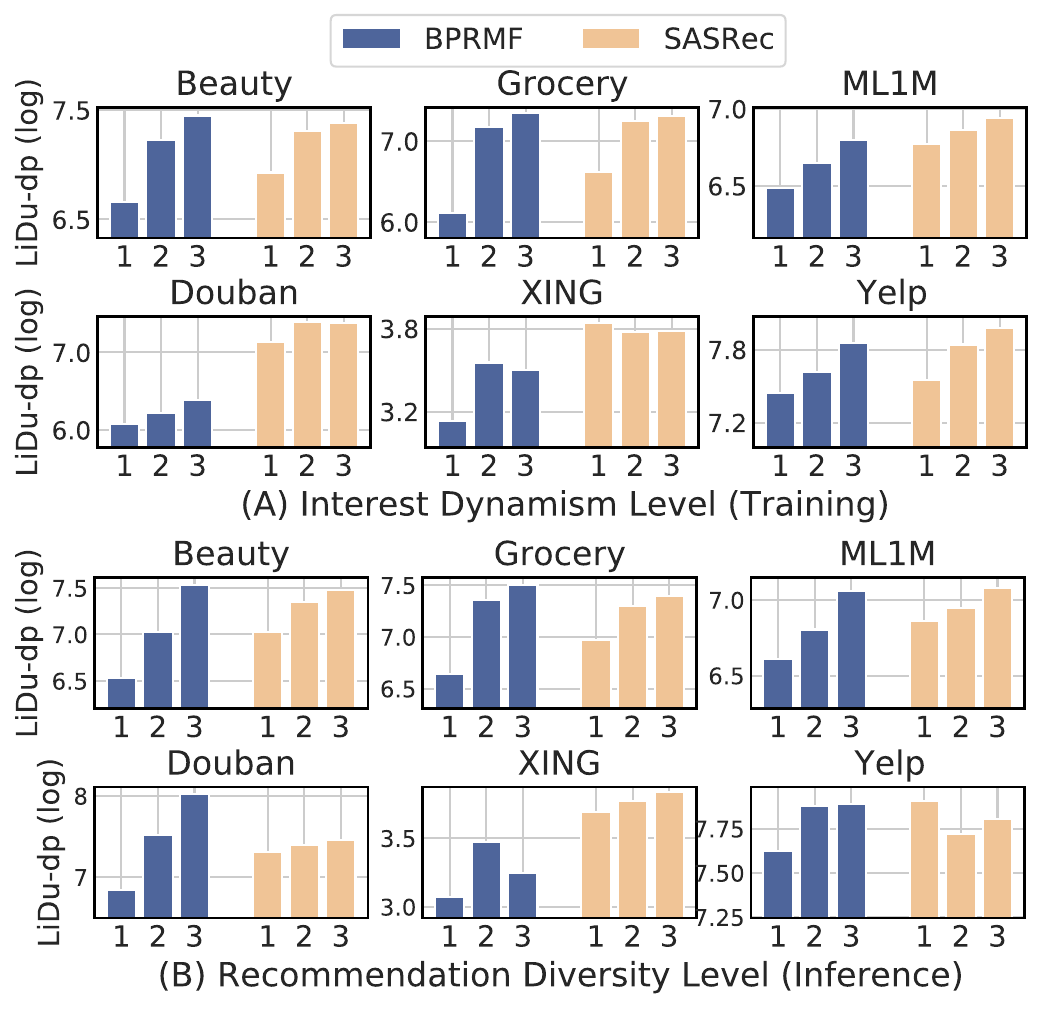}
    \setlength{\abovecaptionskip}{0pt}
    \setlength{\belowcaptionskip}{0pt}
    \caption{Average LiDu-dp of users with (A)~different interest dynamism in the training stage and (B)~different diversity in the inference stage. The LiDu values are log-transformed. Larger dynamic level indicates higher degree of dynamism.}
    \label{fig:exp_property}
    \vspace{-16pt}
\end{figure}

Apart from serving as a performance estimator, LiDu also reflects the model's inner properties.
Here, we analyze LiDu from two perspectives: its relationship with users' history dynamism during training and its association with ranking diversity during inference.
Below, we elaborate on our observations:

\textbf{\textit{Users with more dynamic interests tend to exhibit higher uncertainty as measured by LiDu.}}
We define users' interest dynamism as the average dissimilarities between adjacent history interactions in the training set.
For a user $u$ with $K$ chronological interactions $\mathcal{I}_u=\{i_1,i_2,...,i_K\}$, the dynamism of $u$ is $d_u$:
\vspace{-5pt}
\begin{equation}
    d_u = \frac{1}{K-1}\sum_{k=1}^{K-1}dis(i_k,i_{k+1})
    \label{eq:interest_dynamic}
\end{equation}

where $dis(i_x,i_y)$ is the complement of cosine similarity between item embeddings of $i_x$ and $i_y$.
To visualize the relationship between dynamism and uncertainty, users are grouped into four equal-sized groups based on $d_u$, and the most dynamic group is excluded due to high instability; the average LiDu-dp of the other three groups is shown in Figure~\ref{fig:exp_property}(A).
Due to space limitations, we only show BPRMF and SASRec, and results on other recommenders are included in our repository.
In most experiments, as the dynamism of user interests increases, LiDu-dp increases. 
In particular, the group of users with the lowest dynamism of interests shows a noticeably lower uncertainty. 
Since it is more challenging to learn the preferences of users with more dynamic interests, this reveals that our proposed LiDu can potentially reflect the recommender's degree of learning about different users. 

\textbf{\textit{When generating more diversified lists, recommenders exhibit higher uncertainty as measured by LiDu.}}
In the inference stage, when a recommender has a stronger tendency for exploration, the recommended item lists become more diversified, and thus, the uncertainty should be higher.
Given a recommendation list, $L_u=\{i_1,i_2,...,i_N\}$, for user $u$, the exploration tendency~(i.e., diversity) $e_u$ of $L_u$ is defined as
\begin{equation}
    e_u = \frac{2}{n(n-1)}\sum_{n=1}^{N-1}\sum_{m=n+1}^N dis(i_m,i_n)
\end{equation}
Similar to Figure~\ref{fig:exp_property}(A), we visualize the relation between LiDu-dp and recommendation diversity in Figure~\ref{fig:exp_property}(B). It shows that uncertainties tend to increase when recommendation lists become diverse, which indicates that uncertainty can describe the recommender's confidence in the inference stage. 



%% file: Sections/6.Conclusions.tex
\section{Discussions and Conclusions}

This paper proposes a list-wise uncertainty quantification method, LiDu.
LiDu presents a significant correlation with the recommender's performance on a synthetic dataset and six real-world datasets with a series of recommenders.
Such an indicator can be activated even before recommendations are presented to users, paving the way for developing more robust and interpretable recommendation systems.
Performance estimation without labels is a challenging task, especially considering that user behaviors are unreliable and contain a lot of noise.
However, it is also meaningful and can promote a series of downstream applications as future work:
(1) Data augmentation: In Top-k recommendation, sparse positive samples and ambiguous negative samples are persistent issues~\cite{li2024item}. Uncertainty can serve as an auxiliary annotation for positive and negative samples, turning semi-supervised or even unsupervised ranking tasks into supervised ones.
(2) User-specific adjustment: Uncertainty can identify users for whom the recommendations might be less effective and help adjust the recommendation strategy to mitigate harm to user satisfaction. 
(3) Model selection and ensemble: Uncertainty might enable inter-model comparison after calibration. This is beneficial for choosing a suitable recommender or integrating different recommenders. 
(4) Training optimization: A potential differentiable approximation of uncertainty can be applied to loss functions, optimizing the training process and final performances.

Furthermore, we have revealed the property of LiDu beyond performance accuracy, which is closely related to the recommender's capacity during the training and inference stages. 
This indicates that, apart from serving as a performance estimator, uncertainty can provide additional insights based on the self-awareness of the recommenders.
At this stage, we only focus on user uncertainties. 
Modeling the uncertainty of items may help model cold items and address fairness issues, which is also a valuable exploration direction.